\documentclass{article}
\usepackage{spconf,amsmath,graphicx}
\usepackage{multirow}
\usepackage{url}
\usepackage{hyperref}
\usepackage{float}
\usepackage{amsfonts}
\title{SAN: a robust end-to-end ASR model architecture}

\begin{document}
%\ninept
%
\name{Zeping Min$^{\star1}$ \qquad 
Qian Ge$^{\star1}$ 
\thanks{$^{\star}$Equal contribution} \qquad 
Guanhua Huang$^{\star2}$}
\address{$^{1}$ Peking University \\
$^{2}$ University of Science and Technology of China}

\maketitle
\begin{abstract}
In this paper, we propose a novel Siamese Adversarial Network (SAN) architecture for automatic speech recognition, which aims at solving the difficulty of fuzzy audio recognition. Specifically, SAN constructs two sub-networks to differentiate the audio feature input and then introduces a loss to unify the output distribution of these sub-networks. Adversarial learning enables the network to capture more essential acoustic features and helps the models achieve better performance when encountering fuzzy audio input. We conduct numerical experiments with the SAN model on several datasets for the automatic speech recognition task. All experimental results show that the siamese adversarial nets significantly reduce the character error rate (CER). Specifically, we achieve a new state of art 4.37 CER without language model on the AISHELL-1 dataset, which leads to around 5\% relative CER reduction. To reveal the generality of the siamese adversarial net, we also conduct experiments on the phoneme recognition task, which also shows the superiority of the siamese adversarial network.  
\end{abstract}
\section{Introduction}
\label{sec:intro}

Automatic speech recognition (ASR) is a task with a wide range of application scenarios. There has a long history on automatic speech recognition (ASR). Before the popularity of deep learning, the HMM-GMM \cite{bansal2008improved} models are widely used in the automatic speech recognition community. 
In HMM-GMM architecture, each frame of input corresponds to a label category, and the labels need repeated iterations to ensure more accurate alignment. With the development of deep learning, automatic speech recognition entered the end-to-end era. Recently, researchers have presented many end-to-end deep learning speech recognition methods \cite{watanabe2018espnet, gulati2020conformer, yao2021wenet, baevski2020wav2vec}, which gain better performance and easier training. 

However, these models still suffer from the fuzzy audio problem. For example, the word "wood" and "world" have very similar pronunciations. It may be more difficult to distinguish after adding various noises to the actual scene and hence the model may suffer from them. Most previous works add an additional language model to the decoder to deal with the fuzzy audio problem \cite{yao2021wenet, huang2020class, kannan2018analysis, weiran22_interspeech}. The language model performs well when fuzzy audio is easy to distinguish semantically. For instance, in the audio with transcription of "There are billions of people in the world.", the raw model output may have a high probability on both "There are billions of people in the world." and "There are billions of people in the wood." predictions. And it is easy to eliminate the latter interference option by using the language model.

Unfortunately, this does not work all the time. When the interference option is also semantically meaningful, the language model can not help eliminate the wrong option and may even mislead the output, such as the "I like the wood" and "I like the world.". Both of them are semantically meaningful, but "I like the world" are more common. For input wave with the ground truth transcription "I like the wood.",  the speech recognition model output may also have a high probability on the transcription of "I like the world.". Further, the language model may also vote for the wrong option "I like the world.". Finally, the model is prone to make mistakes on the audio input.

In this paper, inspired by \cite{ganin2016domain}, we propose a novel siamese adversarial net (SAN) architecture for automatic speech recognition, which aims at solving the difficulty of recognizing fuzzy audio. In detail, SAN consists of two weight-shared sub-networks, which employ the dropout layers to mix different noises into acoustic features and make the acoustic features of the two sub-networks different. Then a Kullback–Leibler (KL) divergence is leveraged to minimize the output distributions of these two sub-networks, which boosts the model to learn the essential acoustic features to help the model deal with the fuzzy audio input. As experimental results, we achieve a new state of art 4.37 CER on AISHELL-1 dataset, which leads to around 5\% relative CER reduction to the previous. In summary, our contributions are as follows:
\begin{itemize}
    \item We propose a novel siamese adversarial net (SAN) architecture, solving the difficulty of recognizing fuzzy audio by adversarial learning with two subnets.
    \item We fulfill the gap that few works take care of the fuzzy audio recognition in the acoustic model itself.

    \item We achieve a new state of art 4.37 CER without  language model on the AISHELL-1 dataset, which leads to around 5\% relative CER reduction previous. Besides,  A large number of experiment  results show that our SAN architecture is effective.

\end{itemize}

\section{Related work}

Since the CTC \cite{graves2006connectionist} was proposed, a large number of speech recognition systems based on deep learning have emerged, including \cite{watanabe2018espnet}, \cite{gulati2020conformer}, and
\cite{yao2021wenet}, etc. Although these deep learning models have achieved impressive results in speech recognition systems, they usually make mistakes when encountering fuzzy audio input.

Most previous works combine the language model (LM) and the automatic speech recognition model to alleviate the fuzzy audio problem, such as adding an additional language model \cite{yao2021wenet, huang2020class} when decoding. Some recent works further use the fusion of pre-trained language models \cite{kannan2018analysis}. Moreover, several works introduce the language model for enhancing some sub-tasks in ASR, such as the rare word recognition \cite{weiran22_interspeech}.

In this article, we propose a novel siamese adversarial net (SAN) architecture, which uses dropout to differentiate acoustic features and leverage KL divergence to unify these features. Similar to us, there exist some works in NLP that employs dropout layers to generate positive samples as data augmentation \cite{gao-etal-2021-simcse} or use KL divergence as a loss function to reduce the gap between training and inference \cite{wu2021r}.

\section{Method}
\begin{figure}[!t] 
\centering 
\includegraphics[width=0.5\textwidth]{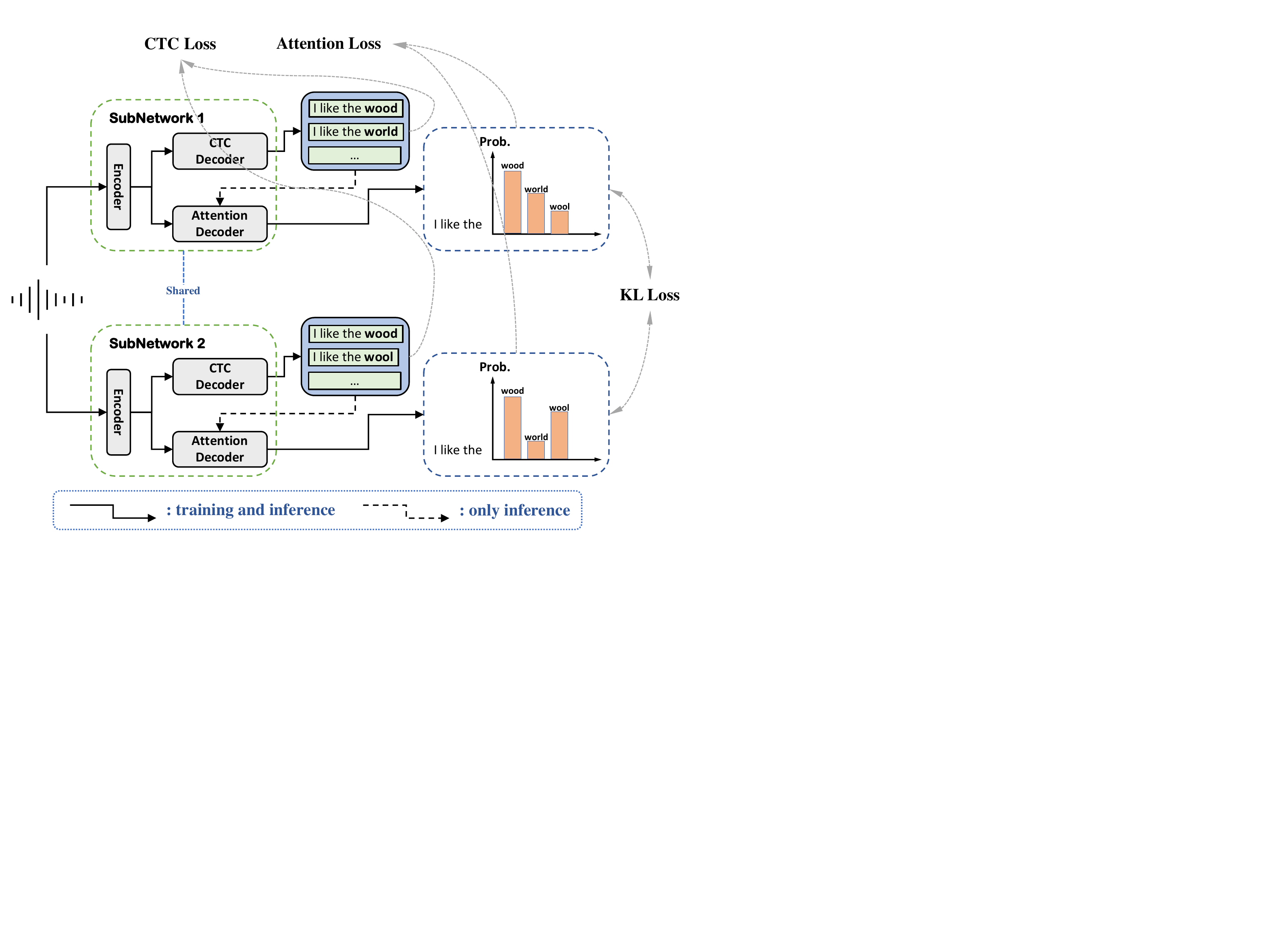} 
\caption{The model architecture of SAN. SAN consists of two sub-networks that share the same parameters. The dropout layers differentiate the acoustic features in two sub-networks, while the KL loss is employed to unify these acoustic features. In the sub-network of SAN, a conformer encoder is used to encode the raw audio sequence, and a CTC decoder and a transformer decoder are leveraged to combine the streaming decoding and the attention decoding.}
\label{framework} 
\end{figure}

As shown in Figure \ref{framework}, SAN first encodes the speech input to obtain the acoustic features (\ref{Encoding}). Then the CTC decoder generates streaming results while the attention decoder generates a global result (\ref{Decoding}). To deal with the fuzzy audio problem, the siamese networks conduct adversarial training to extract the essential feature for predicting the fuzzy part (\ref{Siamese Computing}).

\subsection{Encoding} \label{Encoding}
Given a sequence of acoustic feature 
$S = \{s_1, s_2, ..., s_{N_s}\}$ where 
$s_i \in \mathbb{R}^{d}$, 
$d$ is the embedding dimension and $N_s$ is the length of the speech sequence. 
We encode the acoustic feature by a conformer \cite{gulati2020conformer}:
\begin{equation}
\setlength\abovedisplayskip{3pt}%shrink space
\setlength\belowdisplayskip{3pt}
H = Conformer([s_1, s_2, ..., s_{N_s}])
\end{equation}
and obtain the speech's hidden representation 
$H = [h_1, h_2, ...,$
$ h_{N_s}] \in \mathbb{R}^{N_s \times d}$.

\subsection{Decoding} \label{Decoding}
\subsubsection{CTC Decoding}
After getting the hidden representation $H$ of speech, a CTC decoder is employed for speech decoding, which allows repetitions of labels and occurrences of blank labels. 
\begin{equation}
\setlength\abovedisplayskip{3pt}%shrink space
\setlength\belowdisplayskip{3pt}
P_{CTC} = Softmax(Linear([h_1, h_2, ..., h_{N_s}]))
\end{equation}
where $P_{CTC} = [p_1, p_2, ..., p_{N_s}] \in \mathbb{R}^{N_s \times N_v}$ is the text vocabulary softmax score of speech sequence and $N_v$ is the length of text vocabulary. For model training, we leverage the beam search CTC loss:
\vspace*{-0.5\baselineskip}
\begin{equation} \label{CTC Loss}
\mathcal{L}_{CTC} = BeamCTC([p_1, p_2, ..., p_{N_s}])
\vspace*{-1\baselineskip}
\end{equation}

\subsubsection{Attention Decoding} \label{Rescoring}
Moreover, we employ a transformer decoder to achieve auto-regressive decoding with the help of attention between the acoustic features and text translation results.
\begin{equation}
\setlength\abovedisplayskip{3pt}%shrink space
\setlength\belowdisplayskip{3pt}
[t_1, t_2, ..., t_{n}] = Decoder(t; H)
\end{equation}
where $t_i \in \mathbb{R}^d$ is $i$-th token representation in text translation results, $t \in \mathbb{R}^d$ is the initial query, and $H$ is the acoustic feature. Then we can get the softmax scores and the cross entropy loss for attention translation:
\begin{equation}
\setlength\abovedisplayskip{3pt}%shrink space
\setlength\belowdisplayskip{0pt}
P_{Attn} = Softmax(Linear([t_1, t_2, ..., t_{n}]))
\vspace*{-0.5\baselineskip}
\end{equation}

\begin{equation} \label{Attn Loss}
\setlength\abovedisplayskip{0pt}%shrink space
\setlength\belowdisplayskip{3pt}
\mathcal{L}_{Attn} = CrossEntropy(P_{Attn}, Y)
\end{equation}
where $P_{Attn} = [p_1, p_2, ..., p_{n}] \in \mathbb{R}^{n \times N_v}$ is the softmax score of text results and $Y$ is the ground truth text.

In the model inference, the transformer decoder rescores the text sequence results from CTC decoding instead of directly decoding for acoustic features. In detail, top-k text translation results of CTC decoding are fed to the transformer decoder. Then each text sequence in CTC results obtains a new score which is the product of softmax scores for all tokens. Finally, the text sequence with the best new score is selected to be the final translation result.

\subsection{Siamese Computing} \label{Siamese Computing}
To alleviate the fuzzy audio problem, we extract the essential feature by achieving a siamese adversarial training. 

We build a siamese network \cite{koch2015siamese}, which consists of two sub-encoder-decoder networks sharing the same parameters. The randomness of dropout layers in the two sub-network will differentiate the features of speech, while a KL-divergence loss is minimized to make the distribution of translation results the same. In the adversarial learning of two sub-networks and the KL loss, the randomness of dropout layers mask different part of acoustic features to differentiate the acoustic features, the KL loss aims to learn the same output distribution condition on different fuzzy acoustic features, which promotes the essential feature in the prediction of the fuzzy part.

In detail, The same model input is first fed to the two encoder-decoder modules independently. Then two different output distributions are obtained after the computation of two sub-networks since the dropout layers in these two sub-networks randomly mask the model features, which differentiates the acoustic features.

\vspace*{-1.5\baselineskip}
\begin{align} 
    P_1 & = SubNetwork-1([s_1, s_2, ..., s_{N_s}]) \\
    P_2 & = SubNetwork-2([s_1, s_2, ..., s_{N_s}])
\vspace*{-1.5\baselineskip}
\end{align}

Where $P_1, P_2$ are output distributions of two sub-networks. Finally, a KL-divergence loss is employed to unify the output of two sub-network. 

\vspace*{-1\baselineskip}
\begin{equation} \label{KL Loss}
\mathcal{L}_{KL} = KL-Divergence(p_1, p_2)
\vspace*{-1\baselineskip}
\end{equation}

\subsection{Training}
For training SAN model, we make a weighted sum over three losses in Equation \eqref{CTC Loss}, \eqref{Attn Loss} and \eqref{KL Loss} as follows \footnote{We plus the loss from two sub-networks together for CTC loss and Attention loss.}:
\begin{equation} \label{KL loss}
\setlength\abovedisplayskip{3pt}%shrink space
\setlength\belowdisplayskip{3pt}
\mathcal{L}_{All} = \lambda_1 \mathcal{L}_{CTC} + \lambda_2 \mathcal{L}_{Attn} + \lambda_3 \mathcal{L}_{KL}
\end{equation}
More training details are listed in Section \ref{Settings}.

\section{Experiments}
\subsection{Datasets}
In order to evaluate our proposed SAN model,
we conduct experiments on three datasets: AISHELL-1 \cite{bu2017aishell}, AISHELL-3 \cite{shi2020aishell} and 
AIDATATANG-200zh\footnote{\url{https://openslr.org/62}}.
All these three datasets are open-source Chinese Mandarin speech corpus. AISHELL-1 \cite{bu2017aishell} dataset consists of a 150-hour training set, a 10-hour validation set, and a 5-hour test set. AISHELL-3 \cite{shi2020aishell} contains 85 hours of emotion-neutral recordings spoken by 218 native Chinese mandarin speakers and a total of 88035 utterances. AIDATATANG  contains 200 hours of acoustic data, and the training set, validation set, and testing set are divided into a ratio of 7: 1: 2.

\subsection{Experimental Settings} \label{Settings}

We adopt 12 conformer layers as our encoder. For the decoder, 6 transformer layers are used as our attention decoder for all experiments. Each transformer layer is built with an embedding dimension of 256 and 4 attention heads.

The experiments on AISHELL-3 were trained using two 24Gb memory RTX3090 GPUs with batchsize=32 for each GPU, and we conducted experiments on AISHELL-1 and AIDATATANG with four 16Gb memory P100 GPUs and batchsize=12. We train the models for 240 epochs on AISHELL-1, 65 epochs on AISHELL-3 and 120 epochs on AIDATATANG. We use Adam optimizer with a learning rate of 0.002 during the training process, and a learning rate schedule with 25000 warmup steps on AISHELL-1, 2000 warmup steps on AISHELL-3, and 15000 warmup steps on AIDATATANG. Moreover, we get our final model by averaging the last 30 models on AISHELL-1 and AIDATATANG, and the last 15 models on AISHELL-3. In experiments on all three datasets, the weights of losses are $\lambda_1 = 0.5, \lambda_2 = 0.5, \lambda_3 = 2$.

\subsection{Main Results}
\begin{table}[h]
\scalebox{0.8}{
\begin{tabular}{c|ccc}
\hline
                   & \multicolumn{3}{c}{datasets}                                                            \\ \hline
models             & \multicolumn{1}{c|}{AISHELL-1}     & \multicolumn{1}{c|}{AISHELL-3}     & AIDATATANG    \\ \hline
ESPnet Transformer & \multicolumn{1}{c|}{6.70}          & \multicolumn{1}{c|}{/}             & /             \\
CAT                & \multicolumn{1}{c|}{6.34}          & \multicolumn{1}{c|}{/}             & /             \\
WeNet              & \multicolumn{1}{c|}{4.61}          & \multicolumn{1}{c|}{8.71}          & 4.71          \\
SAN(Ours)          & \multicolumn{1}{c|}{\textbf{4.37}} & \multicolumn{1}{c|}{\textbf{8.68}} & \textbf{4.46} \\ \hline
\end{tabular}
}
\caption{The results of CER (character error rate) on the different test sets under different models. SAN achieves the best performance on all datasets.}
\label{t1}
\end{table}

Table \ref{t1} shows the character error rate (CER) on the test set of three different datasets using End2End architecture. 
We compare our model with ESPnet Transformer \cite{espnet}, CAT \cite{an2020cat} and WeNet \cite{yao2021wenet}. To make the comparison fair, all experiments are conducted without extra language models.
From the results, we can see that our model achieves competitive results of 4.37 CER(character error rate) on AISHELL-1, which outperforms the previous WeNet model and achieves a 5.2\% relative character error rate reduction.
On AIDATATANG, our model also returns state-of-the-art results with 4.46 CER and outperforms the WeNet, and achieves a 5.3\% relative CER reduction. 
On AISHELL-3, our model outperforms all three other models as well.

\subsection{Case Study}
\begin{figure}[htb]
        \centering
        \includegraphics[width=0.4\textwidth]{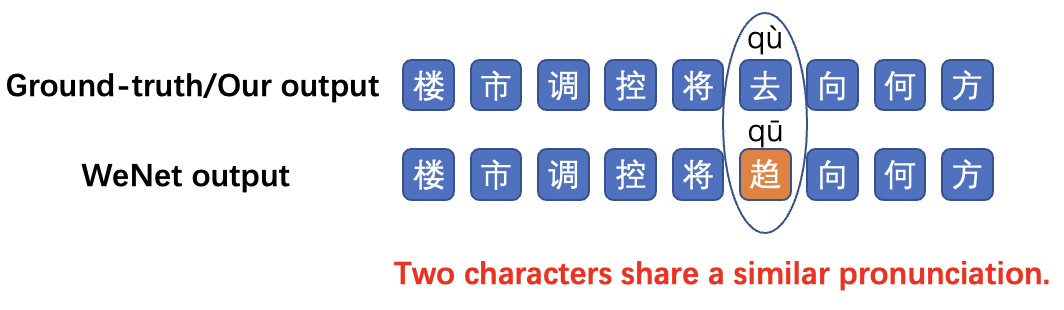}
	\caption{A case of a fuzzy audio problem.}
 \label{case study}
\end{figure}

In this section, we show that our SAN model can mitigate fuzzy speech recognition errors. In Fig. \ref{case study}, we show an intuitive example. The first line is the output text of our model towards audio and is exactly the same as the ground truth. The second line is the inference output from WeNet facing the same input audio.

% We can find that both two models can correctly recognize most characters in the sentence. The only mistake in WeNet's output is replacing a character with another character. The wrong character has a very similar pronunciation to the correct one, they only differ in tone. What's more, both the wrong and correct characters can be combined with the following character to form two commonly used words which have clear meanings. After searching in the training set, we found both the two characters and two words appeared several times. So we claim that our proposed SAN is more discriminative. We speculate it is because the siamese adversarial structure in SAN has the capability of capturing and focusing on the most essential features. 
% If sapce left, change here!!
We can find that our proposed SAN is more discriminative for the fuzzy audio above. We speculate it is because the siamese adversarial structure in SAN helps the model to have the capability of capturing and focusing on the essential features.

\section{Ablation Study}
In order to better understand the SAN architecture, we conduct some ablation experiments.
\subsection{Effective of adversarial architecture}
To show the effectiveness of adversarial architecture in SAN, we compare the experiment results of SAN and SAN without adversarial architecture. In order to be more convincing, we tried the cases where the encoders and subnets in the SAN are transformers and LSTMs, respectively. The result of SAN with transformer encoder (SAN with Transformer)) and LSTM subnet (SAN with LSTM)) are shown in Table \ref{t2} and Table \ref{t3}.
We can clearly see that the model with the siamese adversarial structure achieves a lower CER than the model without it. 
\begin{table}[h]
\scalebox{0.8}{
\begin{tabular}{c|c|c|c}
\hline
models                                                                                      & decode mode & AISHELL-1 & AIDATATANG \\ \hline
\multirow{4}{*}{\begin{tabular}[c]{@{}c@{}}SAN with Transformer\\ without adversarial\end{tabular}}
& attention   & 5.69             & 5.78              \\& greedy      & 5.92 & 6.87 \\& CTC prefix      & 5.91             & 6.85              \\& attention rescore     & 5.30             & 5.62              \\ \hline
\multirow{4}{*}{\begin{tabular}[c]{@{}c@{}}SAN with Transformer\end{tabular}}  
& attention   & \textbf{5.43}    & \textbf{5.22}     \\      & greedy      & \textbf{5.82}    & \textbf{6.62}     \\
& CTC prefix      & \textbf{5.82}    & \textbf{6.61}     \\ & attention rescore     & \textbf{5.05}    & \textbf{5.40}     \\ \hline
\end{tabular}
}
\caption{Performance comparison of SAN (Transformer Encoder) with/without siamese adversarial learning architecture. After using the transformer encoder instead of the conformer encoder, SAN with siamese adversarial architecture still outperforms SAN that without siamese adversarial architecture.}
\label{t2}
\end{table}

\begin{table}[h]
\scalebox{0.8}{
\begin{tabular}{c|c|c}
\hline
models        & CER on dev set                    & CER on test set  \\ \hline
SAN with LSTM without adversarial   & 16.82           & 17.45                     \\
SAN with LSTM & \textbf{15.86}  & \textbf{17.04}  \\ \hline
\end{tabular}
}
\caption{
Performance comparison of SAN(LSTM Encoder) with/without siamese adversarial learning architecture. After replacing the conformer encoder with LSTM, SAN with siamese adversarial architecture still gains lower CER than SAN that without siamese adversarial architecture.}
\label{t3}
\end{table}

% We test on using 12 transformer layers as our encoder, and compare it with the model removing our siamese structure as well as the KL-loss mentioned in \ref{Siamese Computing}.  The result is shown in Table \ref{t2}. We traverse these four decoding methods: attention decoder, greedy decoder, ctc prefix decoder and attention rescoring decoder. From Table \ref{t2}, one can straightforwardly get the following observations.
% \begin{itemize}
%     \item The model with the siamese adversarial structure achieve a lower CER than the model without that.  
%     \item The conformer encoder in our SAN model is better than other possible alternatives (such as transformer).
%     % \item Among all these four decode modes, attention rescoring mode tends to achieve the best performance. 
% \end{itemize}

% \subsection{Replacing Transformer into LSTMs}
% In order to further explore the effectiveness of our proposed siamese network architecture, we replaced the transformer-based encoders and decoders into LSTMs\footnote{We borrow some codes from 
% \url{https://github.com/jackaduma/LAS_Mandarin_PyTorch}.}.
% The results are shown in Table \ref{t3}. The observation is that with the help of siamese network, the LSTM-based encoder/decoder model still achieves a lower CER on both dev set and test set. Furthermore, LSTM tends to own a much worse ability compared to transformers. 

\subsection{Phone recognition Task.}
To demonstrate the effectiveness of our SAN model in other relative scenarios. We conducted experiments on phoneme recognition, which is an easier task. To save computing resources, we simplified the net SAN with LSTM above by removing the CTC-loss and only using cross-entropy as our training loss on Timit dataset \cite{timit}, which is a widely used benchmark for recognition of clean speech with a small vocabulary. The outputs are mapped to 39 phonemes to compute the PER(phone error rate) as metrics.

\begin{table}[H]
\scalebox{0.8}{
\begin{tabular}{c|c|c}
\hline
models        & PER on dev set & PER on test set  \\ \hline
LSTM only     & 23.28   & 24.48       \\
Simplified SAN with LSTM  & \textbf{23.05}  & \textbf{24.14}\\ \hline
\end{tabular}
}
\caption{The results of PER (phone error rate) on Timit under different models. After adding the siamese adversarial learning on the simple LSTM model, the SimSAN significantly reduces the PER on the Timit dataset.}
\label{t4}
\end{table}
The results are shown in Table \ref{t4}. Even the simplified SAN(LSTM) model archives a distinct decrease in PER for both the test set and the dev set. 

\section{Conclusion}
We propose a novel Siamese Adversarial Network (SAN) architecture for automatic speech recognition. To the best of our knowledge, this is the first architecture that specifically focuses on the difficult problem of recognizing fuzzy audio in an acoustic model. The SAN architecture can capture key acoustic features and helps the model achieve better performance when faced with fuzzy audio inputs. Experiments on multiple datasets show that our SAN architecture works well and achieves new sota on the AISHELL-1 dataset, 4.37 CER without language model, leading an around 5\% relative improvement. We hope that our architecture can make a difference in ASR tasks, especially in fuzzy speech recognition scenes.

% method named META-AUDIO to build a mandarin character scale audio database. 
% Also, we raise the CAMP procedure to generate audio from a character scale. Combining these two methods, we tend to generate pseudo-speech data conveniently. Through numerical experiments on several representative datasets, one can obtain competitive CER results using limited real data and our pseudo data, which validates the great effectiveness and low (real) data dependency of our methods. For those languages (e.g. dialects) that are difficult to obtain sufficient audio data, we hope that our methods can make a great contribution.

\bibliographystyle{IEEEbib}
\bibliography{strings,refs}

\end{document}